\documentclass[twocolumn]{aastex6}

\usepackage[english]{babel}

\usepackage[T1]{fontenc}
\usepackage{ae,aecompl}

\usepackage{fancyhdr}
\usepackage{amsfonts}
\usepackage{amsmath}
\usepackage{amssymb}
\usepackage{layout}
\usepackage{graphicx}
\usepackage{multirow}
\usepackage{color}
\usepackage{multirow}

\usepackage{times}
\usepackage{natbib}
\def\be{\begin{equation}}
\def\ee{\end{equation}}

\newif\ifAMStwofonts
\AMStwofontstrue

\graphicspath{{./fig/}}

\shorttitle{Cosmography approach to dark energy cosmologies}
\shortauthors{Rezaei et al.}

\begin{document}

\title{Cosmography approach to dark energy cosmologies: new constrains using the Hubble diagrams of supernovae, quasars and gamma-ray bursts}
\author{ Mehdi Rezaei \altaffilmark{1}}
\author{ Saeed Pour Ojaghi  \altaffilmark{1}}
\author{ Mohammad Malekjani \altaffilmark{1}}

\affil{\altaffilmark{1} Department of Physics, Bu-Ali Sina University, Hamedan 65178, Iran }


\begin{abstract}
	In the context of cosmography approach and using the data of Hubble diagram for supernovae, quasars and gamma-ray bursts, we study some DE parametrizations and also the concordance $\Lambda$CDM universe. Using the different combinations of data sample including ({\it i}) supernovae (Pantheon), ({\it ii}) Pantheon + quasars and ({\it iii}) Pantheon + quasars + gamma-ray bursts and applying the minimization of $\chi^2$ function of distance modulus of data samples in the context of Markov Chain Monte Carlo method, we first obtain the constrained values of the cosmographic parameters in model independent cosmography scenario. We then investigate our analysis, for different concordance $\Lambda$CDM cosmology, $w$CDM, CPL and Pade parametrizations. Comparing the numerical values of the cosmographic parameters obtained for DE scenarios with those of the model independent method, we show that the concordance $\Lambda$CDM model has a serious tension when we involve the quasars and gamma-ray bursts data in our analysis. While the high redshift quasars and gamma-ray bursts can falsify the concordance model, our results of cosmography approach indicate that the other DE parametrizations are still consistent with these observations.  
	
\end{abstract}
\maketitle

\section{Introduction}

Recent advances in observational cosmology have revealed that the current universe has experienced a stage of accelerated expansion.
This expansion can be well explained by introducing an exotic component with negative pressure, dubbed dark energy (DE)which violates the strong energy conditions, $\rho_x +3 p_x > 0$. This expansion can also be justified by modifying the standard theory of gravity on extragalactic scales \citep{Riess1998,Perlmutter1999,Kowalski2008}. In the framework of general relativity (GR), it appears that approximately $70 \%$ of the energy budget of the universe is in the form of dark energy \citep{Bennett:2003bz,Spergel:2003cb,Peiris:2003ff}. The cosmological constant $\Lambda$ in which the equation of state (EoS) parameter is equal to $-1$, is the most likely possibility for dark energy.  Although by assuming the cosmological constant and cold dark matter (CDM) in the context of standard $\Lambda$CDM cosmology, one can serve the purpose the accelerated expansion of the universe and the model is in good agreement with the cosmological observations, it suffers from the serious problems of cosmic coincidence and the fine tuning \citep{Weinberg1989,Padmanabhan2003,Copeland2006}.

Also, from the observational point of view, the $\Lambda$CDM cosmology plagued with some significant tensions in estimation of some key cosmological parameters. In particular, the first tension concerns the discrepancy between the amplitude of matter fluctuations from large scale structure (LSS) data \citep{Macaulay:2013swa}, and the value predicted by CMB experiments based on the $\Lambda$CDM. As the other tension, the Lyman-$\alpha$ forest measurement of the Baryon Acoustic Oscillations (BAO) reported in \citep{Delubac:2014aqe}, suggests a smaller value of the matter density parameter ($\Omega_{\rm m}$) in comparison with the value obtained by CMB data. Furthermore, there is a statistically significant disagreement between the value of Hubble constant measured by the classical distance ladder and that obtained from the Planck CMB data \citep{Freedman:2017yms}. Quantitatively speaking, the $\Lambda$CDM cosmology deduced from Planck CMB data predicts $H_0 = 67.4\pm0.5$ km/s/Mpc \citep{Aghanim:2018eyx}, while from the Cepheid-calibrated SnIa \citep{Riess:2019cxk} we have $H_0=74.03\pm 1.42$ km/s/Mpc. To solve these problems, various kinds of DE models have been proposed in literature \citep{Veneziano1979,Erickson:2001bq,Thomas2002,Armendariz2001,Caldwell2002,Padmanabhan2002,Gasperini2002,Elizalde:2004mq,Gomez-Valent:2014fda}. Comparing with different observations, some of these models have been ruled out and many of them lead to good consistency with data \citep[see also][]{Malekjani:2016edh,Rezaei:2017yyj,Malekjani:2018qcz,Lusso:2019akb,Rezaei:2019roe,Lin:2019htv,Rezaei:2019xwo,Rezaei:2020mrj}. The results of these investigations show that by using the current observations, it is difficult to distinguish different DE models. This confusion about different DE models suggests that a more conservative way to justify the  cosmic acceleration, relying on as less model dependent quantities as possible, is welcome. As a solution, the well known model independent approach which is commonly used in the literature for testing the fitting capability of models with data, is cosmography (see Sect. \ref{sect:cosmography})). Applying the cosmographic approach to distinguish between different DE models was proposed in \citep{Sahni:2002fz,Alam:2003sc}. Using cosmographic parameters in \citep{Capozziello:2009ka}, authors tried to constraint a cosmological model, $f(R)$-gravity. Because these parameters are model-independent, they lead to natural "priors" to any theory. Using cosmography, the authors of \citep{Capozziello:2009ka}  have discussed how $f(R)-$gravity could be useful to solve the problem of mass profile and dynamics of galaxy clusters. In \citep{Capozziello:2011tj}, they studied the possibility to extract the model independent information about the dynamics of the universe by using cosmography approach. Their results showed that in the context of cosmography approach, our predictions considerably deviate from the $\Lambda$CDM cosmology. Based on the cosmography approach, authors of \citep{Capozziello:2018jya} constrained the late time evolution of the Universe using the low-redshift observations. Their results confirmed the tensions with $\Lambda$CDM model at low redshift universe. The authors of \citep{Lusso:2019akb} assumed two different cosmographic models consisting of a fourth-order logarithmic polynomial and a fifth-order linear polynomial, and fitted these models with different data sets. Then, by comparing the results with the expectations from concordance $\Lambda$CDM model, they found significant tensions between the best-fit cosmographic parameters and the concordance $\Lambda$CDM model. The cosmographic approach also is used in \citep{Li:2019qic} to determine the spatial curvature of the Universe. They showed that by combining the supernovae (Pantheon sample), the latest released cosmic chronometers and the BAO measurements, the most favored cosmography model prefers a non-flat universe with $\Omega_K=0.21\pm0.22$.
Following these works, in this paper we want to study some relevant DE models including the standard $\Lambda$CDM, $w$CDM, Chevallier-Polarski-Linder (CPL) and Pade parameterizations in the context of cosmography approach. By combining different data sets including the distance modulus of quasars, the Pantheon and and publicly available gamma-ray burst (GRB) data, we try to find the best-fit values of cosmographic parameters using the minimization of $\chi^2$ function based on the Markov Chain Monte Carlo (MCMC) method. Notice that we first obtain the best fit values of the cosmographic parameters without considering a cosmological model. We then put constraints on the free parameters of the models under study. Using the constrained values and their confidence regions within $1-\sigma$ uncertainties of cosmological parameters of the models, we compute the best fit values of the cosmographic parameters for each model. By comparing the computed cosmographic parameters of the models and those obtained from model independent approach, one can examine the cosmological models against observation.
The layout of our paper is as follows: In Sect. \ref{sect:cosmography}, we present the cosmographic approach. Then we introduce the observational data which we have used in our analysis. In Sect.\ref{sect:models}, we first briefly introduce the DE models and parametrizations in our study and then present the numerical results. In Sect.\ref{sec:numerical} we present discussions based on our numerical results for different models . Finally in Sect.\ref{conlusion}, the paper is concluded.

\section{cosmographic approach}\label{sect:cosmography}

Recently, the cosmographic approach to cosmology commonly used in the literature in order to obtain as much information as possible directly from observations. In this approach without addressing issues such as which model of DE is required to satisfy the accelerated expansion of the Universe, and just by assuming the minimal priors of homogeneity and isotropy we can study the evolution of the Universe.
Cosmography provides information about cosmic flow and it's evolution derived from measured distances, by using Taylor expansions of the basic observables\citep{Demianski:2016dsa}. The distance - redshift relations obtained from these expansions only rely on the assumption of the Friedman-Lemaitre-Robertson-Walker(FLRW) metric and are therefore fully model independent. Firstly, we introduce the cosmographic functions by the first five derivatives of scale factor $a(t)$ as follows \citep{Visser:2003vq}:

\begin{eqnarray}\label{eq1}
{Hubble function:}~~& H(t)=\frac{1}{a}\frac{da}{dt}\;,
\end{eqnarray}

\begin{eqnarray}\label{eq2}
{deceleration function:}~~& q(t)=-\frac{1}{aH^2}\frac{d^2a}{dt^2}\;,
\end{eqnarray}

\begin{eqnarray}\label{eq3}
{jerk function:}~~& j(t)=\frac{1}{aH^3}\frac{d^3a}{dt^3}\;,
\end{eqnarray}

\begin{eqnarray}\label{eq4}
{snap function:}~~& s(t)=\frac{1}{aH^4}\frac{d^4a}{dt^4}\;,
\end{eqnarray}

\begin{eqnarray}\label{eq5}
{lerk function:}~~& l(t)=\frac{1}{aH^5}\frac{d^5a}{dt^5}\;.
\end{eqnarray}

The cosmographic parameters ($H_0,q_0,j_0,s_0 \& l_0$) are corresponding to the present values of the above functions. Furthermore, it is easy to find the relation between the derivatives of the Hubble parameter and the cosmographic parameters as follows:

\begin{eqnarray}\label{h1}
\dot{H}=-H^2(1+q)\;,
\end{eqnarray}

\begin{eqnarray}\label{h2}
\ddot{H}=H^3(3q+j+2)\;,
\end{eqnarray}

\begin{eqnarray}\label{h3}
\dddot{H}=H^4(-3q^2-12q-4j+s-6)\;,
\end{eqnarray}

\begin{eqnarray}\label{h4}
\ddddot{H}=H^5(30q^2+60q+10qj+20j-5s+l+24)\;,
\end{eqnarray}
where each over-dot denotes a derivative with respect to cosmic time $t$. One can compute the Taylor Series expansion of the Hubble parameter to the forth order in redshift $z$ around it's present value $z = 0$

\begin{eqnarray}\label{hexpandz}
H(z)=H\vert_{z=0}+\frac{dH}{dz}\vert_{z=0} \dfrac{z}{1!}+\frac{d^2H}{dz^2}\vert_{z=0} \dfrac{z^2}{2!} + \nonumber\\
\frac{d^3H}{dz^3}\vert_{z=0} \dfrac{z^3}{3!}+\frac{d^4H}{dz^4}\vert_{z=0} \dfrac{z^4}{4!}\;,
\end{eqnarray}
The above Taylor series expansion is valid for small redshifts $z <1$, whereas much of the most interesting recent observational data sets occur at higher redshifts $z>1$. In the other word, the radius of convergence of any series expansion in redshift is equal or less than $(z\lesssim 1)$, and thus any z-based expansion will break down at $z > 1$. To solve this problem we use an improved redshift definition which commonly used in literature, the y-redshift $y=\frac{z}{z+1}$ \citep{Capozziello:2011tj}. Although changing the z-redshift in to the y-redshift will not change the physics, but it can improve the series of convergence. In terms of the y-redshift, we see that the radius of convergence of a Taylor expansion is $(\lesssim 1)$ which correspond to $z\rightarrow\infty$. Thus, using y-redshift definition, we can use the Taylor expansion of the Hubble parameter at any higher redshifts as the following form \citep{Capozziello:2011tj}:

\begin{eqnarray}\label{hexpandy}
H(y)=H\vert_{y=0}+\frac{dH}{dy}\vert_{y=0} \dfrac{y}{1!}+\frac{d^2H}{dy^2}\vert_{y=0} \dfrac{y^2}{2!} + \nonumber\\
\frac{d^3H}{dy^3}\vert_{y=0} \dfrac{y^3}{3!}+\frac{d^4H}{dy^4}\vert_{y=0} \dfrac{y^4}{4!}\;.
\end{eqnarray}

 We note that there are some other procedures which can solve the convergence problem. In \citep{Capozziello:2020ctn}, authors compared some of these procedures to find the best approach to explain low and high redshift data sets. They have expanded the luminosity distance $d_L$, using Taylor series and its alternatives, rational polynomials and auxiliary variables. Their results show that at low redshifts there is no apparent need to adopt the y-variables or rational polynomials instead of Taylor series. But, differences appear at high redshifts, where the results of \citep{Capozziello:2020ctn} indicate that (2,1) polynomial performs better than the y-variables. In this work we use different observations in the redshift range up to $z \sim 7$. Thus we can not use the Taylor expansion and we should apply one of its alternatives. Since we have not using the high redshift CMB data, so we can use an alternative approach having good performance at low and intermediate redshifts. Assuming this condition and in order to prevent the complexity arising from inserting more additional degrees of freedom,  in this work we select the y-redshift procedure.
Now by changing the time derivatives of Eqs.(\ref{h1}-\ref{h4}) in to derivatives with respect to $y$, inserting the results in Eq.(\ref{hexpandy}) and using Eqs.(\ref{eq1}-\ref{eq5}), we will have:

\begin{eqnarray}\label{ey}
E(y)=\dfrac{H(y)}{H\vert_{y=0}}=1+k_1 y+\dfrac{k_2y^2}{2}+\dfrac{k_3y^3}{6}+\dfrac{k_4y^4}{24}\;.
\end{eqnarray}
where different $k_i$ are:

\begin{eqnarray}\label{k1}
k_1=1+q_0\;,
\end{eqnarray}

\begin{eqnarray}\label{k2}
k_2=2-q^2_0+2q_0+j_0\;,
\end{eqnarray}
\begin{eqnarray}\label{k3}
k_3=6+3q^3_0-3q^2_0+6q_0-4q_0j_0+3j_0-s_0\;,
\end{eqnarray}

\begin{eqnarray}\label{k4}
k_4=-15q^4_0+12q^3_0+25q^2_0j_0+7q_0s_0-4j^2_0-\nonumber \\
16q_0j_0-12q^2_0+l_0-4s_0+12j_0+24q_0+24\;.~~
\end{eqnarray}
In the above equations $q_0,j_0,s_0$ and $l_0$ are the current values of cosmographic parameters. By knowing the evolution of $E$ as a function of redshift, we can investigate the evolution of cosmic fluid. In this paper we want to put constraint on the cosmographic parameters using the Hubble diagrams of low redshift observational data. To do this, we set the current value of cosmographic parameters ($q_0,j_0,s_0$ and $l_0$) as the free parameters in MCMC algorithm. Then, the best fit values for the free parameters are those which can minimize the $\chi^2$ function. Notice that the $\chi^2$ function is defined based on the distance modulus of observational objects. The Hubble diagram of low-redshift observational data used in this work is as follows:

\begin{itemize}
\item Pantheon sample: This sample as a set of latest data points for the apparent magnitude of type Ia supernovae (SNIa) \citep{Scolnic:2017caz} in the range $0.01 < z < 2.26$, is one of three sample of data points we use in this work. This sample includes 279 spectroscopically confirmed SNIa discovered by the  Pan-STARRS1(PS1) Medium Deep Survey \citep{Rest:2013mwz,Scolnic:2017caz} in the redshift range $0.03 < z < 0.68$. The pantheon sample also includes the SNIa data from the Sloan Digital Sky Survey (SDSS) \citep{Frieman:2007mr,Sako:2014qmj} and the Supernova Legacy Survey (SNLS) \citep{Conley:2011ku,Sullivan:2011kv}. This sample is the largest combined sample of SNIa data consisting of a total of 1048 data points up to redshift $\sim 2.3$.

	\item		
Gamma-ray bursts (GRBs): 
		The GRBs are the most energetic and powerful explosions in the universe and can be detectable up to very high redshifts. GRBs are the mysterious objects in the universe. A mechanism which indicates the high
		amounts of releasing energy from a typical GRB emits is not yet completely known. Some investigations show that the GRBs are produced by core-collapse events \citep{Meszaros:2006rc}. Despite these difficulties, the GRBs are astrophysical objects for studying the expansion scenario of the universe at high redshifts. In fact, using the Hubble diagrams of GRBs, one can study the expansion rate of the universe and investigate the observational properties of DE up to higher redshifts. One of the most important aspects of the observational property of GRBs is that they show several correlations between spectral and intensity properties (luminosity, radiated energy). \citep{Demianski:2016zxi} proposed an empirical correlation between the observed photon energy of the peak spectral flux, $E_{p,i}$, and the isotropic equivalent radiated energy, $E_{iso}$. This correlation not only provides constraints on the model of the prompt emission, but also naturally suggests that the GRBs can be used as
		distance indicators. In fact to use the GRBs as distance indicators, it is necessary to consistently calibrate this correlation. Unfortunately, due to the lack of GRBs at very low redshifts, the calibration
		of GRBs is more difficult than that of SNIa. In this regard, several calibration procedures have been suggested so far \citep{Dainotti:2008vw,Demianski:2011zj,Demianski:2012ra,Postnikov:2014aua}. Recently, \citep{Demianski:2016zxi} by applying a local regression technique and using the SNIa sample, have constructed a new calibration for the GRB Hubble diagram that can be used for cosmological investigations. They showed that how the $E_{p,i} - E_{iso}$ correlation can be calibrated to standardize the long GRBs and to build a GRB Hubble diagram, which we use to investigate the cosmology at very high redshifts \citep{Demianski:2016zxi}. Notice that their $E_{p,i} - E_{iso}$ correlation has no significant redshift dependence. In this work, we use the 162 data points for distance modulus of GRBs derived and reported in \citep{Demianski:2016zxi}. This sample contains the low and high redshifts GRBs in the range of $ 0.03 < z < 6.67$. More details and discussions about the calibration method and construction of the Hubble diagram of GRBs can be found in \citep{Demianski:2016zxi,Demianski:2016dsa,Amati:2013sca}.

 \item Quasars: 
quasars are extremely luminous active galactic nucleus (AGN), in which a supermassive black hole (SMBH) is surrounded by a gaseous accretion disk. As gas in the disk falls towards the SMBH, energy is released, which can be observed across the electromagnetic spectrum. The observed properties of a quasars depend on  factors such as the mass of the SMBH and the rate of gas accretion. The spectral energy distribution (SED) of quasars shows the significant emission in the optical-UV band $L_{UV}$, the so-called big blue bump (BBB), with a softening at higher energies \citep{Sanders:1989tu,Elvis:1994us,Trammell:2006ex,Shang:2011qm}
. This emission is thought to origin from an optically thick disc surrounding the SMBH. Also, the X-ray photons, $L_X$, are generated by inverse Compton scattering of disc UV photons by a hot electron plasma, the so-called X-ray corona. Notice that the energy loss through X-ray emission may cool down the electron plasma, if there is no efficient energy transfer mechanism from the disc to the corona. However, the physical nature of such a process is still poorly understood. An important observational feature concerning the connection between the UV disc and X-ray corona is provided by the non-linear correlation between the $L_{UV}$ from the disc and $L_X$ from the corona. The non-linear relationship between $L_X$ and $L_{UV}$ as $\log{L_X} = \gamma L_{UV} + \beta$, has been obtained in both optically and X -ray AGN samples with slope parameter $\gamma$ around $0.5 - 0.7$ \citep{Vignali:2002ct,Strateva:2005hu,Steffen:2006px,Just:2007se,Green:2008yx,Lusso:2009nq,Young:2009jc,Young:2009it,Jin:2012ba} representing that optically bright AGN emits relatively less X-rays than optically faint AGN. It has been shown that such relation is independent of redshift and it is very tight \citep{lusso2016tight}. This relation has also been used as a distance indicator to estimate cosmological parameters. Using the $L_X - L_{UV}$ relationship, \cite{Risaliti:2015zla} have constructed the complete sample of quasar Hubble diagram up to $z \sim 6$, which is in excellent agreement with the analogous Hubble diagram for SNIa in the common redshift range (i.e., $z \sim 0.01 - 1.4$). This capability turns quasars into a new class of standard candles \citep{Lusso:2017hgz}. The main quasars sample is composed of 1598 data points in te range $0.04<z<5.1$. In this work instead of the main sample, we use a binned catalog including 25 datapoint from \citep{Risaliti:2015zla,lusso2016tight}. All the details on sample selection, X-ray, and UV flux computation and the analysis of the nonlinear relation, calibration, and a discussion on systematic errors are provided in \citep{Risaliti:2018reu}.

\end{itemize}

Combining Gamma ray bursts and quasars with Pantheon is motivated, because we can probe a redshift range $( 0.03 < z < 6.67)$ better suited for investigating DE than the one covered by Pantheon sample  $(0.01 < z < 2.26)$. Hence, by adding these data samples to Hubble diagram, we have more observational data at higher redshifts. Using these datasets we calculate the $\chi^2$ function of the distance modulus based on the MCMC algorithm to find the best fit values of cosmographic parameters in a model independent cosmology. To run the MCMC algorithm, we select two different sets of the initial values for free parameters. This can guaranty that our results are independent from the initial values of free parameters. For all of the free parameters we choose big $\sigma$ values to ensure that the MCMC can sweep the whole of the parameters space. Using these choices, we have removed the risk of finding a local best fit values in the parameters space. Notice that for both of the initial value sets, we obtained similar posteriors for $q_0$ and $j_0$. But in the case of $s_0$ and $l_0$, the posteriors are slightly different. Thus we have repeated our analysis by setting an initial value for $s_0$ and $l_0$ between two best fit values obtained in previous steps (we presented the initial values of free parameters in Table \ref{tab:inival}).

\begin{table*}
 \centering
 \caption{Two different sets of initial values for MCMC algorithm (left part) and the the final results (right part). 
}
\begin{tabular}{c  c  c c c c c c c c }
\hline \hline
 &  & initial&values& &$\vert$ & &  best fit values&  \\
parameter & $q_0$ & $j_0$ & $s_0$& $l_0$  & $\vert$ &  $q_0$ & $j_0$ & $s_0$& $l_0$\\
\hline 
Set (1) & $-2.0$ & $5.0$ & $-4.0$ & $-5.0$ & $\vert$   & $-0.838_{-0.04}^{+0.06}$ & $2.27^{+0.25}_{-0.33}$ & $-3.8^{+0.67}_{-1.0}$ &  $-5.2^{+2.2}_{-3.0}$ \\
\hline
Set (2)& $2.0$ & $-5.0$ & $4.0$ & $5.0$ &  $\vert$  &$-0.811\pm 0.090$ & $2.51^{+0.24}_{-0.31}$ &$-0.11^{+0.8}_{-1.7}$ &$0.91^{+2.10}_{-3.7}$ \\
\hline
Final Set & $-0.8$ & $2.5$& $-2.0$&$-3.0$ &  $\vert$ &$-0.819\pm 0.065$ &$ 2.21^{+0.37}_{-0.42}$ &$-3.44^{+0.46}_{-1.5}$ & $-3.8^{+8.2}_{-6.2}$ \\

\hline \hline
\end{tabular}\label{tab:inival}
\end{table*}

In order to see the influence of each data sample of quasars and GRB in our analysis, we consider different combinations of data samples as Pantheon, Pantheon+GRB, Pantheon+quasars and Pantheon+GRB+quasars. For all of these combinations, we do our analysis in order to find the best fit values of the free parameters and their $1-$ and $2-\sigma$ uncertainties. Notice that a procedure to chose the proper initial value for each of the free parameters were described above. The results of our analysis are presented in Tab. \ref{tab:bestmi}. For all combinations of data samples, we can see that the deceleration parameter $q_0$ is tightly constrained. The constraints for jerk parameter $j_0$ is approximately tight. However, our analysis can not put the tight constraints on the snap $s_0$ and lerk $l_0$ parameters. We observe that adding the high redshift observational data of quasars and GRB causes to get higher values of $q_0$ and $j_0$. Due to the large values of uncertainties for $s_0$ and $l_0$, we cannot reach to clear conclusion when we compare the results of different combinations. Notice that the $s_0$ and $l_0$ parameters are appearing in the forth and fifth term of the Eq.\ref{ey} as the coefficient of third and forth order of redshift respectively. In these terms, the big error bar of data points leads to very weak constraints on these two parameters.

\begin{table*}
 \centering
 \caption{The best fit values of cosmography parameters and their $1-\sigma$ uncertainties obtained for different combinations of data samples.
}
\begin{tabular}{c  c  c c c}
\hline \hline
Data sample & $q_0$ & $j_0$ & $s_0$& $l_0$ \\
\hline 
Pantheon & $-0.702\pm 0.104$ &$1.60\pm 0.71$ &$-3.54^{+0.38}_{-1.5}$ &$-4.9^{+6.3}_{-5.0}
$ \\
\hline
Pantheon+GRB &$-0.775\pm 0.048$ &$2.61^{+0.29}_{-0.19}$ &$ 2.8\pm 1.4$ &$-1.3^{+3.0}_{-3.6}$ \\
\hline
Pantheon+quasars  &$-0.844\pm 0.048$ &$ 2.42\pm 0.25$ &$-2.5^{+1.4}_{-1.2}$ &$-3.2^{+2.5}_{-2.1}$  \\
\hline
Pantheon+GRB+quasars  &$-0.819\pm 0.065$ &$ 2.21^{+0.37}_{-0.42}$ &$-3.44^{+0.46}_{-1.5}$ & $-3.8^{+8.2}_{-6.2}$ \\
\hline \hline
\end{tabular}\label{tab:bestmi}
\end{table*}

\section{DE models and parameterizations}\label{sect:models}

In this section we first briefly introduce some DE models and parameterizations which we want to study in cosmography approach. Notice that we also consider the standard $\Lambda$CDM cosmology as a concordance model. Then, by using the data samples presented in previous section and by applying the minimization of $\chi^2$ function based on the MCMC algorithm, we find the best fit values of the cosmological parameters of DE models.
Using the chain obtained for cosmological parameters of each model within $1-\sigma$ level, we compute the best fit and $1-\sigma$ uncertainty of cosmographic parameters for each model. Finally, we will compare the best fit cosmographic parameters of each model with those of the model independent approach obtained in Table (\ref{tab:bestmi}). The DE models that we examine in our analysis are:

\begin{enumerate}
	
\item $wCDM$: The first model is the DE model with constant equation of state (EoS) parameter $w_{de}$. The Hubble parameter of the model in a flat FRW universe reads\citep{Mota:2003tm,Barger:2006vc}:

\begin{eqnarray}\label{wcd}
E^2(z)=\Omega_{m,0}(1+z)^{3}+(1-\Omega_{m0})(1+z)^{3(1+w_{de})}\;,
\end{eqnarray}
where $\Omega_{m,0}$ is the energy density of pressure-less matter at the present time. Notice that we study the model in late time cosmology where the energy density of radiation is negligible.
Using the above equation and re-writing Eqs.(\ref{eq1}-\ref{eq5})in term of redshift, we can obtain cosmographic parameters in the context of $w$CDM cosmology as follows:

\begin{eqnarray}\label{qwcd}
q(z)=\dfrac{\Omega_{m,0}(1+z)^{3}+(1+3w)\Omega_{d,0}(1+z)^{3(1+w)}}{2 E^2(z)}\;,
\end{eqnarray}

\begin{eqnarray}\label{jwcd}
j(z)=1+\dfrac{9w(1+w)\Omega_{d,0}(1+z)^{3(1+w)}}{2 E^2(z)}\;,
\end{eqnarray}

In order to obtain the best fit values and the confidence regions of the cosmographic parameters, we need to obtain the best fit and also the confidence regions of the cosmological parameters $\Omega_{m,0}$ and $w_{de}$ of the model. Notice that in the flat FRW universe, we have $\Omega_{d,0} = 1-\Omega_{m,0}$.
So using the different combinations of observational data: Pantheon, Pantheon + GRB, Pantheon + quasaras and finally Pantheon+GRB+quasars, we obtain the best best fit values of $\Omega_{m,0}$ and $w_{de}$ as well as their confidence regions in $1-\sigma$ uncertainty. Our results are reported in the left part of Table (\ref{bestwcdm}). Using Eqs. (\ref{qwcd}, \ref{jwcd}) and the data of $\Omega_{m,0}$ and $w_{de}$ in $1\sigma$ error, we put constraints on the cosmographic parameters in $wCDM$ cosmology. Results for the best fit values and $1-\sigma$ confidence regions are presented in the right part of Table (\ref{bestwcdm}).

\begin{table*}
 \centering
 \caption{The best fit values of cosmological parameters  in $wCDM$ DE parametrization obtained from the minimization of $\chi^2$ function based on MCMC algorithm (left part) and the best fit values of the cosmographic parameters computed in $wCDM$ model (right part).
}
\begin{tabular}{c  c  c c c  c c}
\hline \hline
 &  &best fit parameters &$\vert$ & & computed & values  \\
Data & $\Omega_{m,0}$ & $w$ & $\vert$ & $q_0$&    & $j_0$\\
\hline 
Pantheon & $0.349^{+0.037}_{-0.029}$ & $-1.25^{+0.15}_{-0.13}$ & $\vert$ & $-0.709^{+0.086}_{-0.075}$ &    & $1.92^{+0.45}_{-0.66}$\\
\hline
Pantheon+GRB & $ 0.352^{+0.036}_{-0.027}$ & $ -1.26\pm 0.14$ & $\vert$ & $-0.714\pm 0.081$ &    & $1.97^{+0.49}_{-0.61}$\\
\hline
Pantheon+quasars & $0.389^{+0.027}_{-0.023}$ & $-1.42^{+0.15}_{-0.13}$ & $\vert$ & $-0.801^{+0.088}_{-0.077}$ &   & $2.69^{+0.53}_{-0.74}$\\
\hline
Pantheon+GRB+quasars & $0.388^{+0.028}_{-0.022}$ & $-1.42^{+0.15}_{-0.13}$ & $\vert$ & $-0.798^{+0.086}_{-0.077}$ &  & $2.66^{+0.53}_{-0.72}$\\
\hline \hline
\end{tabular}\label{bestwcdm}
\end{table*}

\item Concordance $\Lambda$CDM: In fact when we do an analysis on a given DE model, we should redo our analysis for standard $\Lambda$CDM model as a concordance model. So in this part we study the standard model from the viewpoint of cosmography approach. In order to obtain the cosmographic parameters for $\Lambda$CDM model, we can easily set $w_{de}=-1$ in Eqs.(\ref{wcd}-\ref{jwcd}). Then we follow the procedure implemented for $w$CDM model to find the cosmographic parameters in $\Lambda$CDM cosmology. Our results are presented in Table (\ref{bestlcdm}). Notice that in the $\Lambda$CDM model, the jerk parameter is exactly equal to one independent of the values of cosmological parameters.
\begin{table*}
 \centering
 \caption{The best fit values of cosmological parameters obtained for $\Lambda$CDM universe (left part) and the the best fit of cosmographic parameters of the model (right part).
}
\begin{tabular}{c   c c c c  c}
\hline \hline
 &  best fit parameters & $\vert$ & & computed&values  \\
Data  & $\Omega_{m,0}$  & $\vert$ & $q_0$&  &  $j_0$\\
\hline 
Pantheon & $0.285\pm 0.013$ & $\vert$ & $-0.572\pm 0.019$ &  & $1.0$\\
\hline
Pantheon+GRB & $ 0.285\pm 0.012$ & $\vert$ & $-0.572\pm 0.019$ &  & $1.0$\\
\hline
Pantheon+quasars & $0.294\pm 0.012$ & $\vert$ & $-0.559\pm 0.019$ &  & $1.0$\\
\hline
Pantheon+GRB+quasars & $0.294\pm 0.012$ & $\vert$ & $-0.559\pm 0.019$ &  & $1.0$\\
\hline \hline
\end{tabular}\label{bestlcdm}
\end{table*}

\item Pade parametrization: As a well known parametrization for the EoS of DE, we consider the Pade parametrization in this work. The Pade Parametrization is the rational approximation of order $(m,n)$ for an arbitrary function $f(z)$ as follows:

\begin{eqnarray}
f(x) = \frac{a_0 + a_1x + a_2x^2 + ... + a_nx^n}{b_0 + b_1x + b_2x^2 + ... + b_mx^m}\;,
\end{eqnarray} 
where exponents $(m, n)$ are positive and the coefficients $(a_i,b_i)$ are constants \citep{pade1892}. In this work, we consider the Pade expansion of the Eos parameter $w_{de}(a)$
up to the order $(1,1)$ around the variable $(1-a)$, where $a$ is scale factor. Previously, this parametrization has been studied in \citep{Rezaei:2017yyj} in the light of different observational data. But, here we investigate this parametrization from the cosmography point of view. The EoS parameter for the Pade $(1,1)$ parametrization can easily be written as follows\citep{Rezaei:2017yyj,Rezaei:2019hvb}: 
	\begin{eqnarray}\label{pade}
	w_d(z)=\dfrac{w_0+(w_0+w_1)z}{1+z+w_2z}\;.
	\end{eqnarray}

Following  \citep{Rezaei:2017yyj,Rezaei:2019hvb}, we can find the evolution of dimensionless Hubble parameter of Pade parametrization, $E(z)$, as
	\begin{eqnarray}\label{epade}
	E^2(z)=\Omega_{m,0}(1+z)^3+( 1+w_2-\dfrac{w_2}{1+z}) ^{p_1}\times \nonumber \\
	(1-\Omega_{m,0})(1+z)^{p_2}\;,~~~~~~~~
	\end{eqnarray}
	
	where $p_1$ and $p_2$ are:
	
	\begin{eqnarray}\label{epadep1}
	p_1=-3(\dfrac{w_1-w_0w_2}{w_2(1+w_2)}) \;, \nonumber\\
	p_2=3\dfrac{1+w_0+w_1+w_2}{1+w_2} \;.
	\end{eqnarray}
	Using Eq. (\ref{epade}) in Eqs.(\ref{eq1}-\ref{eq3}), we can obtain the cosmographic parameters for Pade the parametrization as follows:
	
	\begin{eqnarray}\label{qpade}
	q(z)=\dfrac{3\Omega_{m,0}(1+z)^3+(1+z)(A_1B_1+C_1D_1)}{2E^2(z)}-1\;,
	\end{eqnarray}

	\begin{eqnarray}\label{qpade2}
	j(z)=1+(1+z)^2\dfrac{2A_1C_1+B_1F_1+G_1D_1}{2E^2(z)}-\nonumber \\
	(1+z)\dfrac{A_1B_1+C_1D_1}{E^2(z)}\;,~~~~~~~~
	\end{eqnarray}
	where constants $A_1, B_1, C_1, D_1, F_1$ and $G_1$ are, respectively, given by:
	
	\begin{eqnarray}\label{qpadea}
	A_1=\dfrac{w_2 p_1}{(1+z)^2}(1+w_2-\dfrac{w_2}{1+z})^{-1+p_1}\;, \nonumber\\
	B_1=\Omega_{d,0}(1+z)^{p_2}\;, \nonumber\\
		C_1=p_2\Omega_{d,0}(1+z)^{-1+p_2}\;, \nonumber\\
		D_1=(1+w_2-\dfrac{w_2}{1+z})^{p_1}\;, \nonumber\\
		F_1=p_1(1+w_2-\dfrac{w_2}{1+z})^{p_1-1}-\dfrac{2p_1w_2}{(1+z)^3}+\nonumber \\
		\dfrac{(p^2_1-p_1)w^2_2}{(1+z)^4}(1+w_2-\dfrac{w_2}{1+z})^{p_1-2}\;, \nonumber\\
			G_1=p_2(p_2-1)\Omega_{d,0}(1+z)^{-2+p_2}\;.
	\end{eqnarray}
	
In a cosmology based on the Pade parametrization for EoS parameter of DE, we have four free parameters including $\Omega_{m,0}$, $w_0$, $w_1$ and $w_2$. We redo our analysis for Pade parametrization like what was done for $w$CDM and $\Lambda$CDM cosmologies. So, we first find the best fit as well as the confidence regions of the parameters within $1-\sigma$ level. Then, we obtain the best fit and the error bar of the cosmographic parameters for Pade approximation for different combinations of data samples. Results are presented in Table (\ref{bestpade}). 
	
	\begin{table*}
		\centering
		\caption{The best fit values of cosmological parameters for Pade parameterization (left part) and the best fit of cosmographic parameters (right part).
		}
		\begin{tabular}{c  c  c c c c c c  c}
			\hline \hline
			& & best fit&parameters & & $\vert$ & & computed&values \\
			Data  & $\Omega_{m,0}$ & $w_0$ & $w_1$ & $w_2$ &$\vert$ & $q_0$&  & $j_0$\\
			\hline 
			Pantheon & $0.330^{+0.060}_{-0.045}$ & $-1.24^{+0.15}_{-0.13}$ &$0.21^{+0.72}_{-0.40}$&$0.16^{+0.66}_{-0.42}$& $\vert$ & $-0.741\pm 0.097$ &  & $2.4\pm 1.0$\\
			\hline
			Pantheon+GRB & $0.327^{+0.063}_{-0.045}$ & $-1.22^{+0.15}_{-0.12}$ &$0.08\pm 0.59$&$0.21^{+0.64}_{-0.40}$& $\vert$ & $-0.725\pm 0.094$ &  & $2.22^{+0.91}_{-1.1}$\\
			\hline
			Pantheon+quasars & $0.402^{+0.033}_{-0.024}$ & $-1.40^{+0.15}_{-0.12}$ &$-0.098^{+0.58}_{-0.74}$&$-0.36^{+0.21}_{-0.63}$& $\vert$ & $-0.756^{+0.11}_{-0.098}$ & &  $2.05^{+0.94}_{-1.3}$\\
			\hline
			Pantheon+GRB+quasars & $0.391^{+0.038}_{-0.026}$ & $-1.41^{+0.15}_{-0.13}$ &$-0.10^{+0.55}_{-0.67}$&$-0.07^{+0.55}_{-0.61}$& $\vert$ & $-0.78\pm 0.10$ &  & $2.5^{+1.0}_{-1.3}$\\
			\hline \hline
		\end{tabular}\label{bestpade}
	\end{table*}

\item CPL parametrization: The other parametrization that we study in this work is the well-known Chevallier-Polarski-Linder (CPL) parametrization of DE in which the EoS parameter is simply expanded around $(1-a)$ by Taylor approximation up to first order, e.g., $w=w_0+w_1z/(1+z)$ \citep{Chevallier2001,Linder2003}. It is easy to see that for a particular value of $w_2=0$, we can recover the CPL parametrization from Pade formula. In the CPL parametrization, the Hubble parameter is written as  \citep{Chevallier2001,Linder2003}:

\begin{eqnarray}\label{ecpl}
E^2(z)=\Omega_{m,0}(1+z)^3+(1-\Omega_{m,0})(1+z)^{3(1+w_0+w_1)}\times \nonumber \\
\exp [-3w_1\dfrac{z}{1+z}]\;.~~~~~~~~
\end{eqnarray}

Hence, inserting Eq. (\ref{ecpl}) into Eqs. (\ref{eq1}-\ref{eq3}), the cosmographic parameters in CPL cosmology are obtained as follows:

\begin{eqnarray}\label{qcpl}
q(z)=\dfrac{A_2[1+z+3(1+z)w_0+3zw_1]+(1+z)B_2}{2(1+z)[A_2+B_2]}\;,
\end{eqnarray}

\begin{eqnarray}\label{jcpl}
j(z)=\dfrac{A_2 C_2+2(1+z)^2 B_2}{2(1+z)^2[A_2+B_2]}\;,
\end{eqnarray}
where the constants $A_2, B_2$ and $C_2$ are:
\begin{eqnarray}\label{aa}
A_2=\Omega_{d,0}(1+z)^{3(w_0+w_1)}\;, \nonumber\\
B_2=\Omega_{m,0}\exp [3 w_1\dfrac{z}{1+z}]\;, \nonumber\\
C_2=9z^2w^2_1+3w_1(1+z)(6w_0z+3z+1)+ \nonumber \\
(1+z)^2(9w^2_0+9w_0+2)\;.
\end{eqnarray}

In this case we have three free parameters including $\Omega_{m,0}$, $w_0$ and $w_1$. The best fit values, the $1-\sigma$ confidence region of these parameters and also the best fit values of the cosmographic parameters of the model are reported in Table (\ref{bestcpl}).

\end{enumerate}

\begin{table*}
 \centering
 \caption{The best fit values of cosmological parameters for CPL parametrization (left part) and the best fit of cosmographic parameters of the model (right part).
}
\begin{tabular}{c  c  c  c c c c c}
\hline \hline
 & & best fit parameters & & $\vert$ & & computed&values  \\
Data  & $\Omega_{m,0}$ & $w_0$ & $w_1$  &$\vert$ &$q_0$&  &$j_0$ \\
\hline 
Pantheon & $0.281^{+0.12}_{-0.059}$ & $-1.17\pm 0.17$ &$0.55^{+1.1}_{-0.52}$& $\vert$ & $-0.74\pm 0.10$ &  & $2.33^{+1.2}_{-0.91}$\\
\hline
Pantheon+GRB & $0.326^{+0.061}_{-0.033}$ & $-1.22\pm 0.15$ &$0.30^{+0.53}_{-0.41}$& $\vert$ & $-0.724^{+0.086}_{-0.075}$ &   & $2.17^{+0.59}_{-0.74}$\\
\hline
Pantheon+quasars & $0.382^{+0.035}_{-0.024}$ & $-1.41\pm 0.14$ &$0.08^{+0.60}_{-0.51}$& $\vert$ &$-0.798\pm 0.090$ & &$2.70\pm 0.85$\\
\hline
Pantheon+GRB+quasars & $0.384^{+0.033}_{-0.022}$ & $-1.41\pm 0.14$ &$0.05\pm 0.50$& $\vert$ & $-0.801\pm 0.090$ &  & $2.71^{+0.82}_{-0.91}$\\
\hline \hline
\end{tabular}\label{bestcpl}
\end{table*}

\begin{figure*} 
	\centering
	\includegraphics[width=15cm]{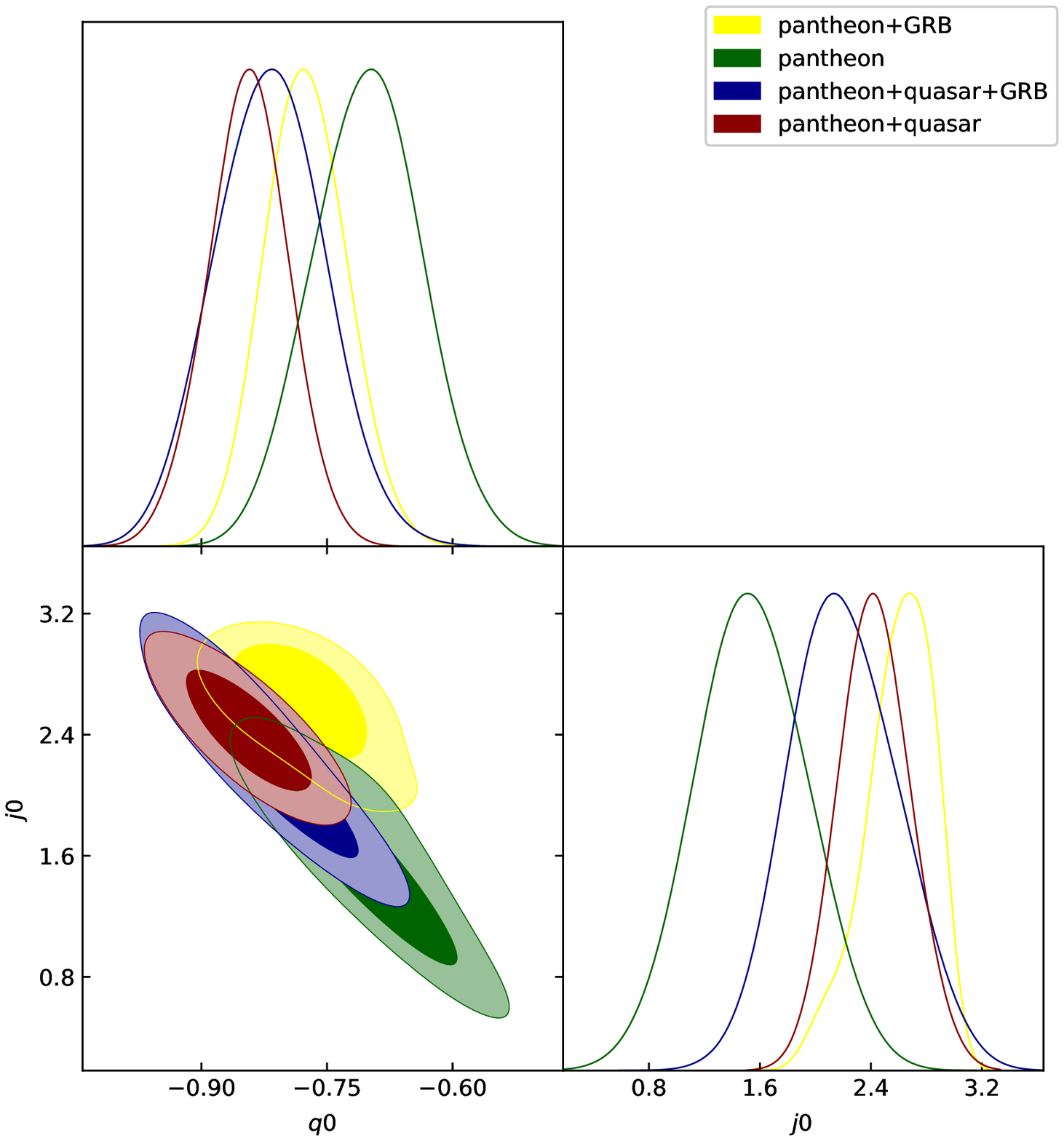}
		\caption{The confidence regions in $q_0 - j_0$ plan obtained in model independent approach using different combinations of Pantheon, quasars and GRB data samples.}
	\label{fig:back}
\end{figure*}

Using these numerical results, in the next section we will compare the cosmographic parameters of the above DE parametrizations with the cosmographic parameters obtained from model independent approach presented in Table (\ref{tab:bestmi}). In Fig. (\ref{fig:back}), we plot the $1-$ and $2-\sigma$ confidence regions of the cosmographic parameters $q_0$ and $j_0$ obtained for model independent approach. We can easily observe that the $2-\sigma$ confidence region of the jerk parameter $j_0$ for different combinations of Pantheon+GRB, Pantheon+quasars and Pantheon+GRB+quasars is above the critical value $j_0=1$. While the confidence region of $j_0$ for Pantheon sample covers the critical point $j_0 = 1$. Hence as a quick result, we can see that the $\Lambda$CDM cosmology has a significant tension with the high redshift GRB and quasars observations. 

\section{discussions}\label{sec:numerical}

\begin{figure*} 
	\centering
	\includegraphics[width=8.5cm]{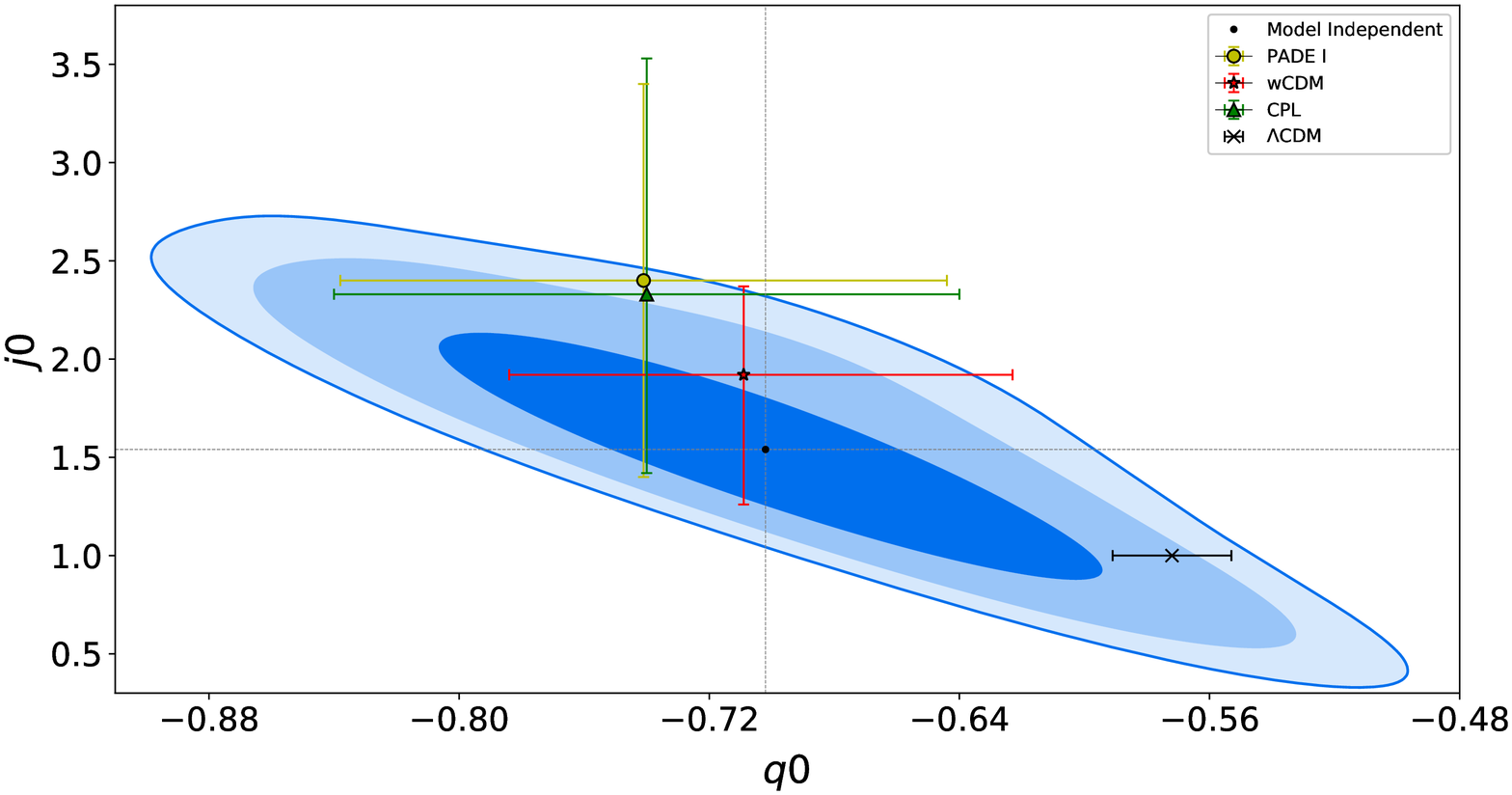}
	\includegraphics[width=8.5cm]{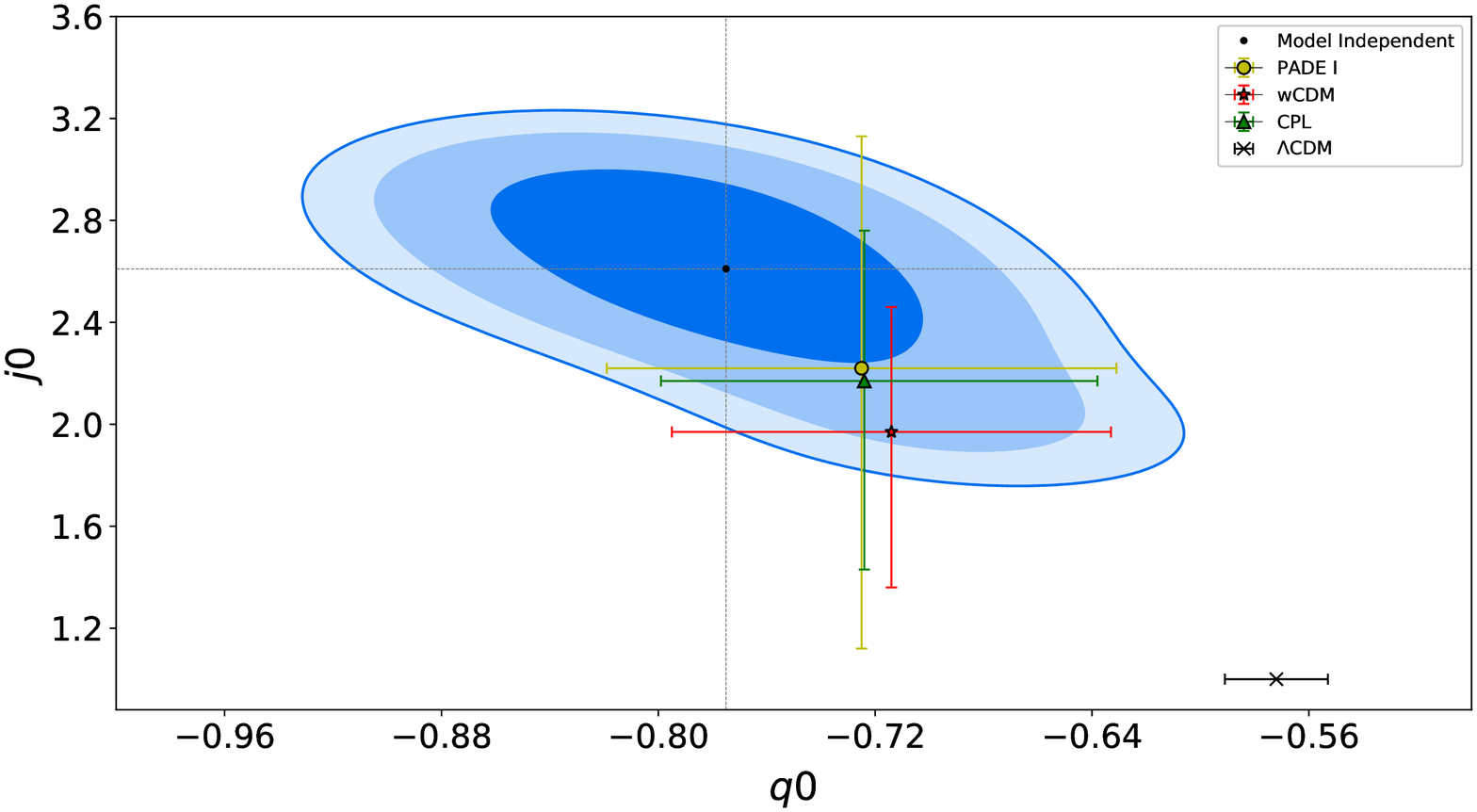}
	\includegraphics[width=8.5cm]{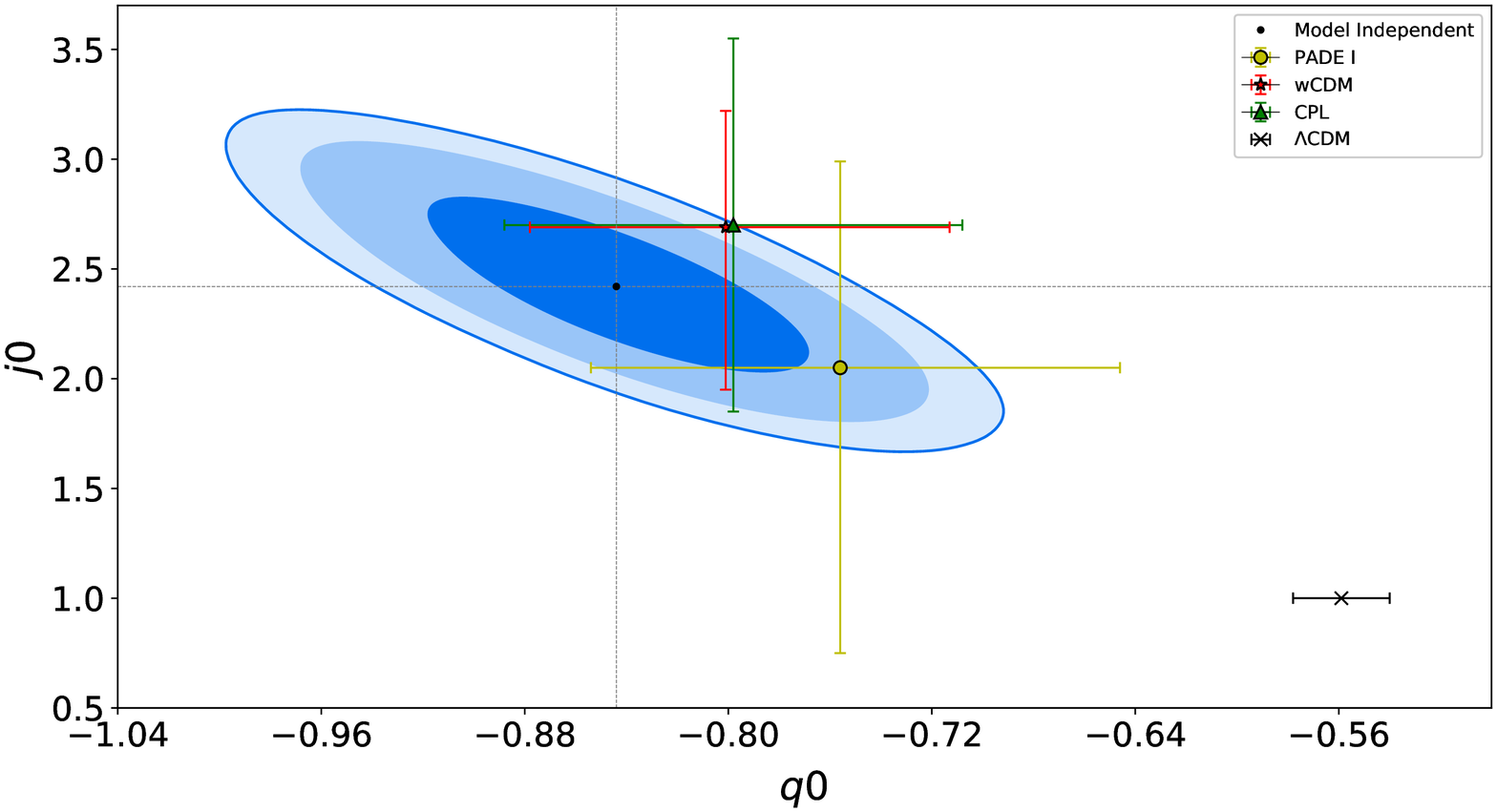}
	\includegraphics[width=8.5cm]{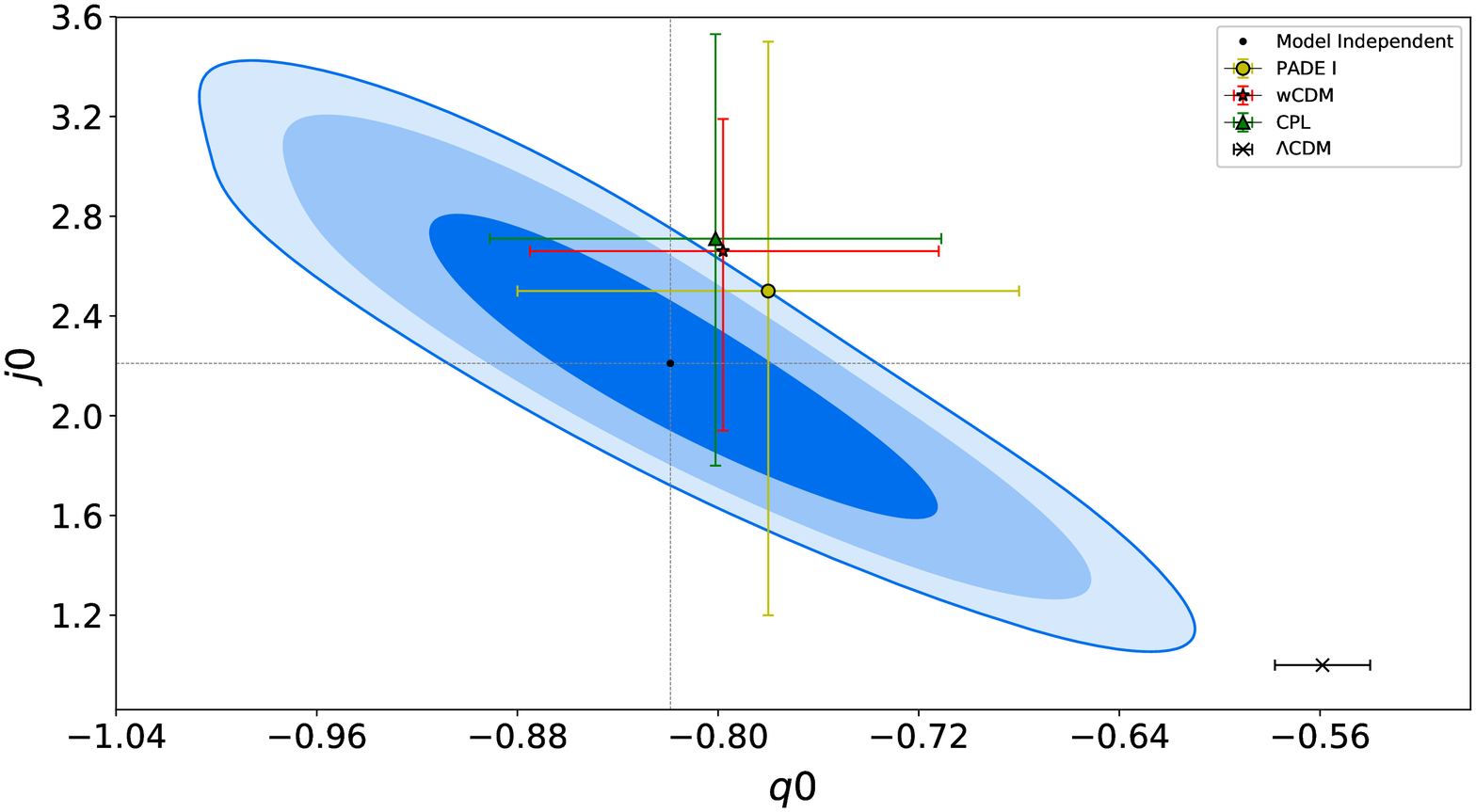}
	\caption{ $3-\sigma$ confidence levels of cosmographic parameters $q_0$ and $j_0$ obtained from model independent approach. Also the best fit values of same parameters with their error bar for different DE scenarios have been shown. The up-left(up-right) panel shows the results obtained using Pantheon (Pantheon+GRB) sample. The bottom-left(bottom-right) panel shows the results obtained using Pantheon+quasars (Pantheon+GRB+quasars) sample.}
	\label{fig:miandmodels}
\end{figure*}

In this section we will compare the numerical results of our analysis obtained in previous sections. As one can see in Tab.\ref{tab:bestmi} our analysis leads to fairly tight constraints on two of the cosmographic parameters, $q_0$ and $j_0$, while the other two parameters of cosmography, $s_0$ and $l_0$, have not been tightly constrained. Furthermore, the best fit values of $s_0$ and $l_0$ are significantly varying based on then the initial conditions of MCMC algorithm. Also their related confidence regions are also so big. Therefore, in order to compare DE models, we just focus on our results for $q_0$ and $j_0$ and ignore the other ones. Notice that the impact of $q_0$ and $j_0$ on the Taylor expansion of the Hubble parameter is much bigger than $s_0$ and $l_0$. Hence in overall, our consideration to compare DE parametrizations based on $q_0$ and $j_0$ can not restrict our conclusion. Based on the results of Table (\ref{tab:bestmi}), when we just use the Pantheon sample, the deceleration parameter $q_0$ has the largest value,$q_0=-0.702$, while adding other data samples to Pantheon leads to smaller value for $q_0$. Thus we can say that the larger value of deceleration parameter $q_0$ is favored by low redshift data points, while using relatively higher redshift data points causes the smaller $q_0$. Oppositely, in the case of jerk parameter we obtain smaller value of $j_0$ when we use the Pantheon sample and larger value of $j_0$ when we add the other data samples to Pantheon. These results are completely in agreement with those of \citep{Lusso:2019akb} which have used one of our combination of data samples (Pantheon+GRB+quasars) in a different way to constraint cosmographic parameters. Our results also nearly confirm the results of \citep{Li:2019qic} which were obtained using different data samples and different approach. Now we compare the best fit values of the cosmographic parameters $q_0$ and $j_0$ for each DE parametrization obtained in previous section with those of the model independent way. Our comparison for different combinations of data samples is described as follows.
\begin{enumerate}
\item {\bf Pantheon sample}: Using the Pantheon sample, The best fit values of deceleration and jerk parameters within $1\sigma$ uncertainty are $q_0=-0.702\pm 0.104$ and $j_0=1.60\pm0.71$ for model independent approach. Our results for $w$CDM model (Table\ref{bestwcdm}) show that both of $q_0$ and $j_0$ are in full agreement (at $1-\sigma$ confidence level) with those we obtained from model independent constrains. In the case of $\Lambda$CDM model, the results are almost different from those of model independent case. For $\Lambda$CDM we have $q_0=-0.572\pm0.019$ ( $j_0=1.0$) which is in $1.25\sigma$ ($0.85\sigma$) tension with the result of $q_0$ ($j_0$) which we have obtained for model independent case (see Tables \ref{tab:bestmi} \& \ref{bestlcdm}). Both of Pade and CPL parametrizations have nearly the same results. The results of $q_0$ in these two parametrizations is in full agreement with the value of $q_0$ which we have found for model independent case, while the value of $j_0$ that we obtained for these parameterizations is in more than $1\sigma$ tension with the $j_0$ of model independent case (see Tables \ref{tab:bestmi}, \ref{bestpade} and \ref{bestcpl}). In Summary, our results are presented in the top-left panel of Fig.\ref{fig:miandmodels} in which the contour plot shows the model independent constraints on $j_0-q_0$ plan up to $3-\sigma$ confidence level and the error bars have used to show the computed value of cosmographic parameters for different cosmological models up to $1-\sigma$. As a result of this part, we can say that using the solely Pantheon sample, the constrained parameters $q_0$ and $j_0$ for $w$CDM, Pade and CPL parametrizations are compatible with model independent constraints in $\sim 1\sigma$ error. While the standard $\Lambda$CDM model can be falsified by $1 \sigma$ uncertainty because of its jerk parameter. Notice that the $\Lambda$CDM model is still consistent with model independent results in $2\sigma$ level.

\item {\bf Pantheon + GRB data} Using this sample the best fit value of cosmographic parameters and their $1-\sigma$ confidence levels in a model independent approach are $q_0=-0.755\pm0.048$ and $j_0=2.61^{+0.29}_{-0.19}$. We see that adding the GRB to Pantheon data leads to smaller value of deceleration parameter and larger jerk parameter compare to solely Pantheon sample. Same as previous part, the $q_0$ parameter in $w$CDM is in full agreement with model independent result. But the $j_0$ parameter in this model has a $\sim3\sigma$ tension with the that of the model independent scenario. Notice that here $1\sigma$ is the average of error bar obtained in model independent $j_0$ (see Table \ref{tab:bestmi}. The results of $\Lambda$CDM model are disappointing in this part. The best fit value of $q_0$ in this model leads to a $3.8\sigma$ tension with that of the model independent one and it's $j_0$ is in  more than $5\sigma$ tension with the bests of the model independent approach. We emphasize that for all comparisons, we define the agreement or tension between DE models and the model independent approach, based on the average of error bar obtained for cosmographic parameters in model independent way presented in Table (\ref{tab:bestmi}.)  
 Like previous part, the results of $q_0$ parameter in the Pade and CPL parametrizations are completely compatible with those of the model independent approach, while the $j_0$ is in a $\sim2\sigma$ tension with the best value of $j_0$ in the model independent approach. Therefore we can say that using the combination of  GRB and Pantheon data points, Pade and CPL  are the best models and $\Lambda$CDM is completely dis-favorable. In the up-right panel of Fig.\ref{fig:miandmodels}, the contour plot shows the best fit of cosmographic parameters and related confidence levels up to $3-\sigma$, obtained using Pantheon+GRB data points in the model independent approach. The best fit values and their error bars obtained for different DE models also are plotted for comparison. We can see that in $q_0 - j_0$ plan, $w$CDM, CPL and Pade parametrizations are located inside the confidence regions while the standard $\Lambda$CDM model is in outside. 

\item {\bf Pantheon + quasars:} Now let see the effect of adding quasars data to Pantheon sample in our analysis. By combination of quasars and Pantheon data sets, the model independent approach leads to $q_0=-0.844\pm0.048$ and $j_0=2.42\pm0.25$. Now we compare this result with the best fit values of $q_0$ and $j_0$ obtained for different DE models. In the case of $w$CDM (see Table\ref{bestwcdm}), we observe that $q_0$ ($j_0$) of the model deviates from model independent values as $0.98\sigma$ ($1\sigma$) region. This result for CPL parametrization (see Table\ref{bestcpl}) is $1\sigma$ and  $1.2\sigma$ deviation, respectively, for $q_0$ and $j_0$. In the case of Pade parametrization (see Table\ref{bestpade}), we obtain $1.9\sigma$ ($1.6\sigma$) deviation from model independent constraints of $q_0$ ($j_0$). Finally in the case of concordance $\Lambda$CDM (see Table\ref{bestlcdm}), we see the big tension between the values of cosmographic parameters of the model and those of model independent approach. Numerically, this tension is approximately $6\sigma$ for both $q_0$ and $j_0$. In the down-right panel of Fig.\ref{fig:miandmodels}, the contour plots show the confidence levels of cosmographic parameters up to $3-\sigma$, obtained using Pantheon+quasars data points in the model independent approach. The best fit values and their error bars of cosmographic parameters obtained for different DE models also are plotted for comparison. We see that the $\Lambda$CDM model is completely outside of the confidence regions, while $w$CDM, CPL and Pade parametrizations are still inside the regions.

\item {\bf Pantheon + quasars + GRB:} In the last step, we combine all of our data samples and compare the results of model independent approach with those of DE parametrizations. As one can see in Tab.\ref{tab:bestmi}, the best fit values of cosmographic parameters in model independent approach are $q_0=-0.819\pm0.065$ and $j_0=2.21_{-0.042}^{+0.37}$. Assuming the results obtained for our models (see Tables \ref{bestwcdm}, \ref{bestlcdm}, \ref{bestcpl}, \ref{bestpade}), we observe that the best fit values of $q_0$ and $j_0$ for $w$CDM model are respectively in $0.3\sigma$ and $1\sigma$ tension with the results of model independent approach. These tensions in the case of $\Lambda$CDM enhance to $3.7\sigma$ for $q_0$ and $4\sigma$ for $j_0$. For Pade parametrization the differences are smaller. Here we have tensions about $0.6-\sigma$ for $q_0$ and $0.8\sigma$ for $j_0$.In the CPL case, $q_0$ parameter has $0.3\sigma$ tension and $j_0$ has $1.3\sigma$ tension with the best fit values in  model independent analysis. 
In the bottom-right panel of Fig.\ref{fig:miandmodels}, the contour plots show the confidence levels of cosmographic parameters obtained using Pantheon+GRB+quasars data points in a model independent approach. The best fit values and their error bars obtained for different DE parametrizations also plotted for comparison. Same as previous parts, the $\Lambda$CDM cosmology are far from the confidence region in $q_0 - j_0$ space, while other models are still not refuted. 
\end{enumerate}      

\begin{figure*} 
	\centering
	\includegraphics[width=8.5cm]{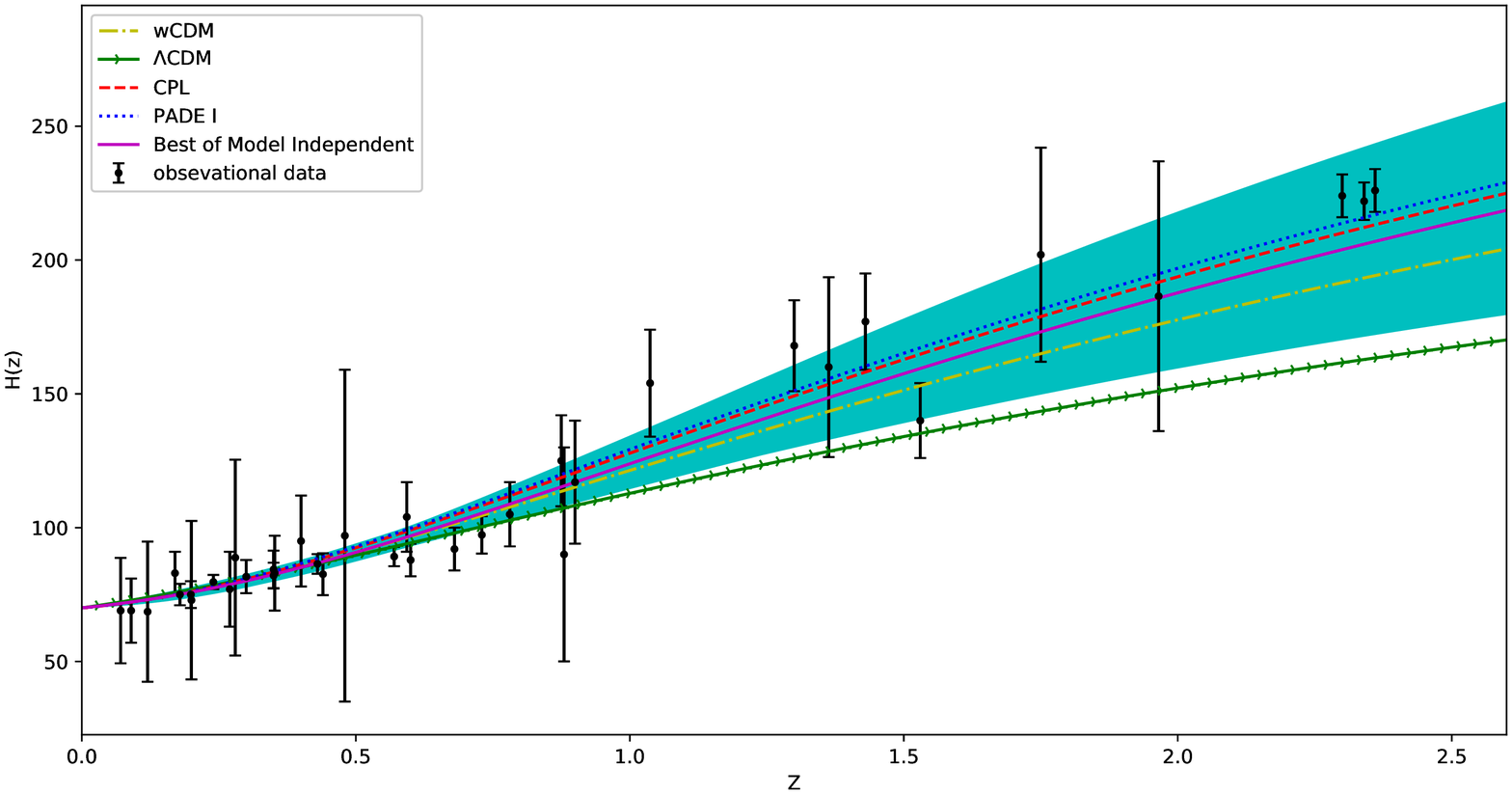}
	\includegraphics[width=8.5cm]{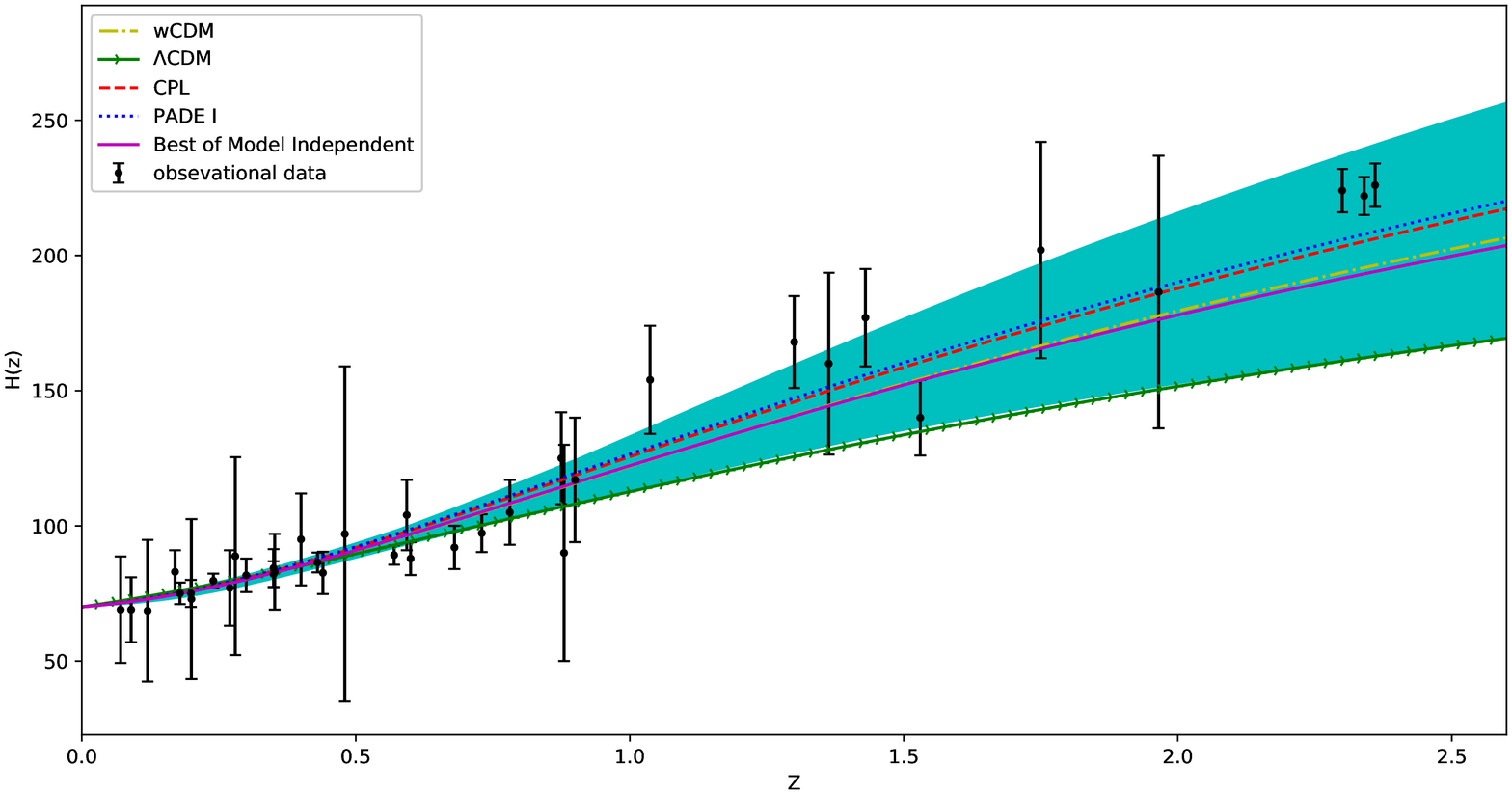}
	\includegraphics[width=8.5cm]{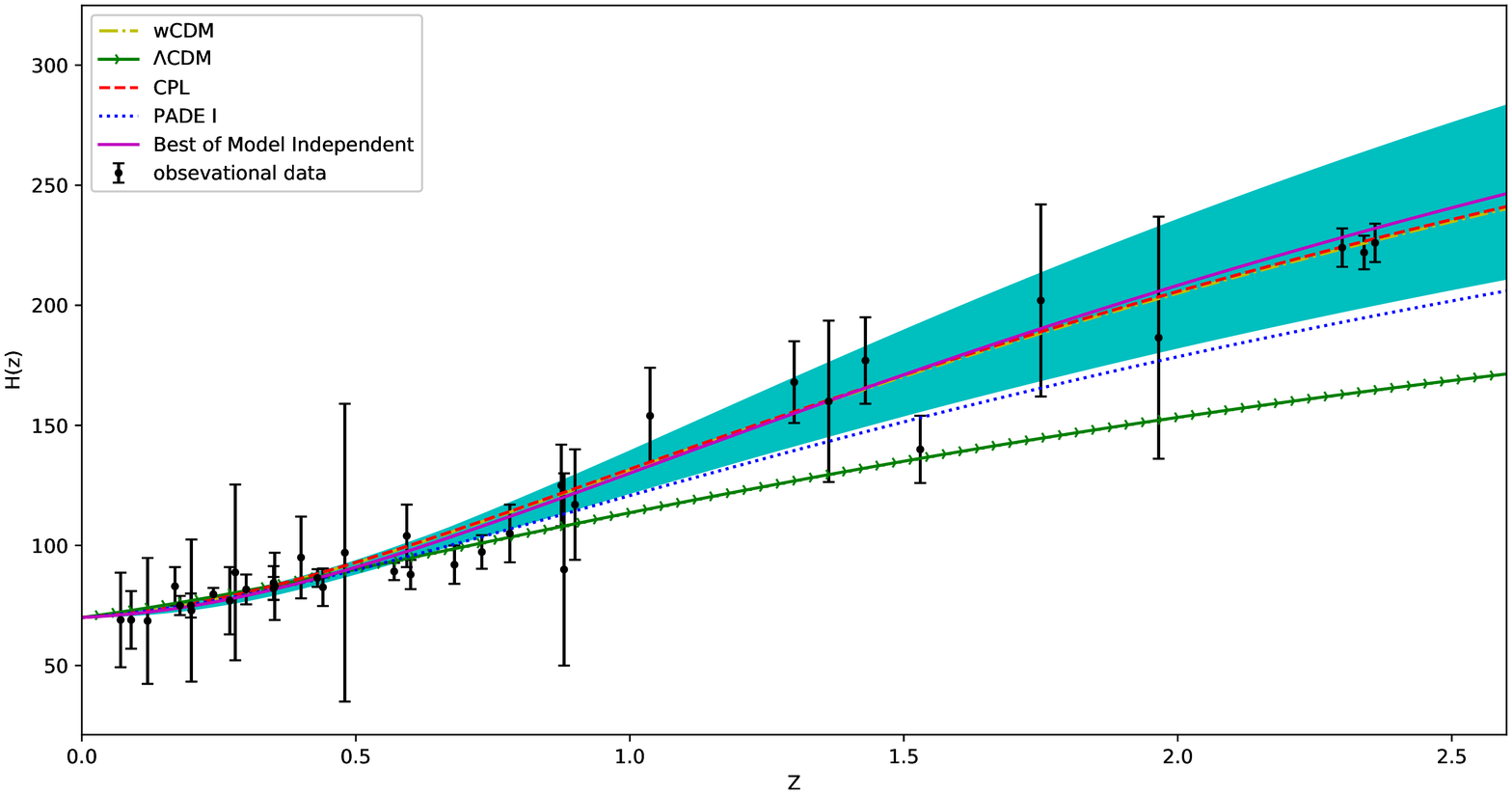}
	\includegraphics[width=8.5cm]{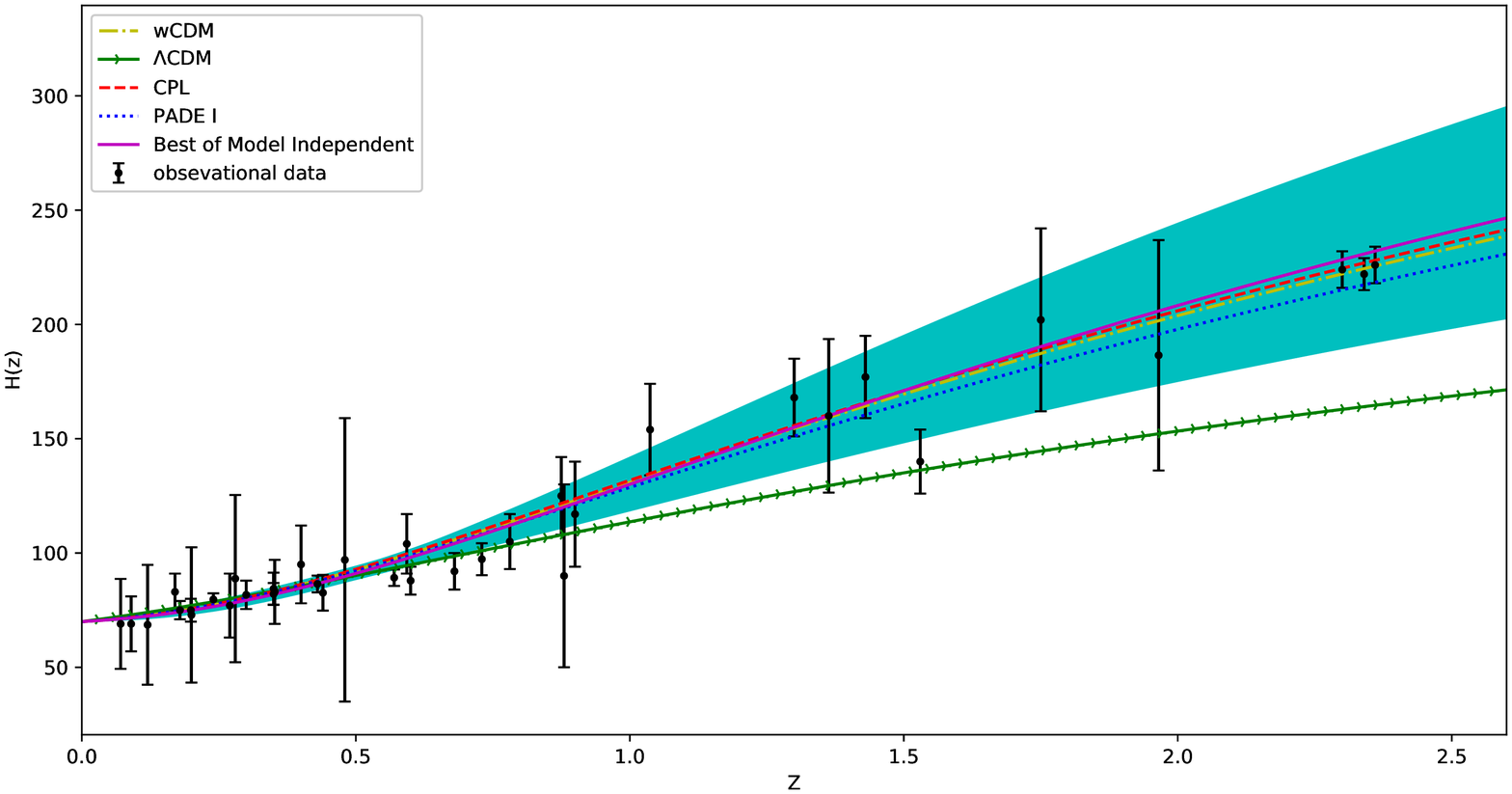}
	\caption{ The reconstructed Hubble parameter $H(z)$ based on the best fit values of the cosmographic parameters $q_0$ and $j_0$ for different model independent approach and DE scenarios studied in this work. The green band shows the $1-\sigma$ confidence region of reconstructed Hubble parameter in model independent method. The up-left(up-right) panel shows the results obtained using Pantheon (Pantheon+GRB) sample. The bottom-left(bottom-right) panel shows the results obtained using Pantheon+quasars (Pantheon+GRB+quasars) sample.}
	\label{fig:hmodels}
\end{figure*}

Now we examine the DE parametrizations and also concordance $\Lambda$CDM universe by reconstructing the Hubble parameter in the context of cosmography approach.
In Fig.\ref{fig:hmodels}, we have reconstructed the redshift evolution of Hubble parameter, $H(z)$, within $1-\sigma$ confidence region, using Eqs.(\ref{ey} - \ref{k4}). Notice that we consider Eq.\ref{ey} up to $y^2$ which involves $q_0$ and $j_0$ parameters. So we can use the best fit values of cosmographic parameters $q_0$ and $j_0$ for model independent approach in Table (Table\ref{tab:bestmi}) and also for DE models and parametrizations in Tables (\ref{bestwcdm} - \ref{bestcpl}). Each of the panels of the figure obtained from one of our combinations of data samples. In all cases, we set $H_0=70$ km/s/Mpc \citep{Abbott:2017xzu}. The  $1-\sigma$ confidence level of $H(z)$ (green band) is calculated by using the upper and lower limits of best fit values of $q_0$ and $j_0$ obtained in model independent approach from Tab.\ref{tab:bestmi}. In the up-left panel we show the reconstructed $H(z)$ obtained from Pantheon sample. The evolution of $H(z)$ for different DE models and parametrizations also plotted for comparison. As one can see in this panel, $H(z)$ curve of $\Lambda$CDM deviates from $1-\sigma$ region at redshifts higher than $z\sim0.8$. While $H(z)$ for other DE parametrizations  evolve within $1-\sigma$ region even at high redshifts. The results obtained from Pantheon+GRB sample are presented in the up-right panel. In this plot same as the previous one, the $H(z)$ curve of $\Lambda$CDM has the maximum differences from the best curve among different models. The bottom-left panel which shows the reconstruction of $H(z)$ for Pantheon+quasars sample, represents the $\Lambda$CDM cosmology as the most incompatible model again. We see that the deviation from confidence region is so big at higher redshifts. Furthermore in this plot, the Pade parametrization also evolves outside of $1-\sigma$ region. Finally in the bottom-right panel, we present the results obtained using Pantheon+GRB+quasars sample. This plot confirms the results of previous panels again. We see that the reconstructed Hubble parameters of $\Lambda$CDM cosmology evolves outside of confidence region at redshifts bigger than $z\sim 0.8$. So among cosmological DE scenarios studied in this work, the $\Lambda$CDM is the worst one.          

\section{Conclusions} \label{conlusion}

In this work we first used the data points of low-redshifts Hubble diagrams for Pantheons, quasars and GRB's to put constraints on the present value of cosmographic parameters in a independent cosmography approach. To do this, we used different combinations of data samples including Pantheon, Pantheon + quasars, Pantheon + GRB and finally Pantheon + quasars + GRB. In the context of cosmography approach, we obtained
the best fit values of cosmographic parameters as well as their confidence regions up to $3-\sigma$ uncertainties for different combinations of data samples. Our results showed that the best fit value of deceleration parameter $q_0$ varies in the range of $-0.844$ to $-0.702$ and the best fit of jerk parameter $j_0$ varies in the range of $1.60$ to $2.61$ for different combinations of data samples. Notice that here we used the Hubble diagrams of quasars and GRB, respectively, derived in \citep{lusso2016tight} and \citep{Demianski:2016zxi}. In the calibration procedure to form the Hubble diagrams of both quasars and GRBs, they have used the SNIa data at low redshifts. Their results for quasars and GRBs samples are consistent with that of the SNIa samples at low redshift universe. Hence we adopted their calibrations and used their Hubble diagrams for quasaras and GRBs. In the case of concordance $\Lambda$CDM cosmology, our results are also compatible with recent work in \cite{Lusso:2019akb}. They confirmed the presence of a tension
between $\Lambda$ cosmology and the best-fit cosmographic parameters $\sim4\sigma$ with SnIa+quasars, at $\sim2\sigma$ with SnIa+GRBs, and at $\>4\sigma$ with the whole SnIa+quasars+GRB data set \cite{Lusso:2019akb}. Furthermore, we studied some relevant DE parametrizations as well as the concordance $\Lambda$CDM cosmology using the  Hubble diagrams of Pantheons, quasars and GRB observations in the context of cosmography approach. The DE parametrizations studied in our analysis are $w$CDM, CPL and Pade parametrizations. Firstly, by using the different combinations of data samples and in the context of MCMC algorithm, we calculate the $\chi^{2}$ function of the distance modulus to find the best fit values  and also the $1-\sigma$ uncertainty of cosmological parameters for each DE parametrization. Using the chain of data obtained for cosmological parameters, we found the best fit values and $1-\sigma$ confidence region of the cosmographic parameters of DE parametrizations. Comparing the results for DE models with those of obtained for model independent approach leads to conclude that which of the model is in better (worse) agreement with Hubble diagrams of Pantheons, quasars and GRB's. In the first stage, 
using the solely Pantheon sample, we found that the $w$CDM model is the most compatible model with the result of model independent constraints and on the other hand the concordance $\Lambda$CDM model is the worst model. In the second step, by combining the GRB data to the Pantheon sample, we obtained disappointed results for $\Lambda$CDM model. In this case $q_0$ parameter of the $\Lambda$CDM has a $3.8-\sigma$ tension with that of the model independent cosmography approach. Moreover, the $j_0$ parameter of $\Lambda$CDM cosmology, has roughly $5-\sigma$ tension with that of the model independent approach. These results will be more frustrated when we see the results of other DE models and parametrizations that we studied in this work. We observed that $w$CDM, CPL and Pade parametrizations are in better agreement with the results of model independent cosmography approach rather than concordance model. In the third and fourth steps, by using the combinations Pantheon+quasars and Pantheon+GRB+quasars data points, we obtained the same results again, supporting our results in previous steps.
So we conclude that the concordance $\Lambda$CDM cosmology has a big tension with the observations of quasars and GRB at higher redhsift. Notice that the DE parametrizations studied in this work sre in better agreement with Hubble diagrams of high redshift quasars and GRB observations. Finally, we reconstructed the Hubble parameter by using the best fit value of cosmographic parameters for both model independent approach, $\Lambda$CDM model and DE parametrizations. We observed that for different data sample combinations, the evolution of reconstructed  $H(z)$ in concordance $\Lambda$CDM model has the maximum deviation from the confidence region compare to different DE parametrizations. Upon this result, we can conclude that among different cosmological models studied in this work, the $\Lambda$CDM has the minimum compatibility with the predictions of model independent approach and thus it is falsified by cosmography approach. The big value of tensions (between $3\sigma$ to $6\sigma$ for different data combinations) that we observed between the cosmographic parameters of $\Lambda$CDM and those we obtained in model independent approach support this claim again that we should explore other alternatives for standard $\Lambda$CDM cosmology. We observed that other DE parametrizations in this study can not be refuted in the context of cosmography approach.
Our results for $\Lambda$CDM cosmology are in agreement with the results of recent work in \citep{Yang:2019vgk,Khadka:2019njj} which was obtained by using a different approach and different data sets. Although, in the literature it has been thoroughly affirmed that the $\Lambda$CDM well describes the evolution of the universe until recent times, but our conclusion confirms the result of \citep{Benetti:2019gmo} representing some big tensions emerge at higher redshifts for $\Lambda$CDM. Our analysis can be extended by calling the other cosmic observations in the context of cosmography approach.

\section{Acknowledgements}
The work of MR has been supported financially by Iran Science Elites Federation.

\bibliographystyle{aasjournal}
\bibliography{ref}
\label{lastpage}

\end{document}